# Generation of Optical Vortex Light Beams by Volume Holograms with Embedded Phase Singularity

A.Ya. Bekshaev*, S.V. Sviridova, A.Yu. Popov, A.V. Tyurin

*I.I. Mechnikov National University, Dvorianska 2, 65082, Odessa, Ukraine*

**Abstract**

Special features of the optical-vortex (OV) beams generated by thick holographic elements (HE) with embedded phase singularity are considered theoretically. The volume HE structure is based on the 3D pattern of interference between an OV beam and a standard reference wave with regular wavefront. The incident beam diffraction is described within the framework of a linear single-scattering model in which the volume HE is represented by a set of parallel thin layers with the "fork" holographic structure. An explicit integral expression is derived for the complex amplitude distribution of the diffracted paraxial beam with OV. The numerical analysis demonstrates that the HE thickness may essentially influence not only selectivity and efficiency of the OV beam generation but also the amplitude and phase profile of the diffracted beam as well as regularities of its propagation. We have studied the generated OV morphology and laws of its evolution; in particular, the possibility of obtaining a circularly symmetric OV beam regardless of the diffraction angle is revealed.

## 1. Introduction

Paraxial light beams with optical vortices (OV) attract growing attention of the physical optics community [1–6]. Special physical features of such beams (wavefront singularities, isolated points and/or compact regions of zero intensity, transverse energy circulation, mechanical orbital angular momentum with respect to the propagation axis, etc.) are interesting for the fundamental physics and promise impressive applications. In particular, the OV beams are suitable for capturing and manipulation of microparticles [7–9], measurement of small displacements and exact localization of optical inhomogeneities [10–13] as well as for improved resolution of spatially overlapping optical signals [14–17] and in schemes of encoding and processing information [18–26]. Being fully understandable in terms of classical optics, beams with OV appear to be unique objects for analyses of fine details of the quantum behavior, including the quantum entanglement and verification of the Bell inequalities [21–26].

Among known approaches to generation and analysis of the OV beams, the most suitable and efficient ones are based on the holographic principle [27–29]. In such techniques, an incident beam interacts with the hologram where the pattern of interference between an OV beam and a certain reference wave with regular structure (a plane or a spherical wave, or a usual laser-generated Gaussian beam) is recorded, and the desirable output beam is formed in some diffraction order. The special feature of such patterns is the fringe bifurcation ("fork" structure) [18,21,26,31–40] that represents embedded phase singularity (EPS) associated with the hologram. In practice, it is not necessary to record the hologram in a real interference process; the pattern can be calculated and fabricated artificially, so that a so-called computer-generated hologram (CGH) with EPS is formed.

---

* Corresponding author. Tel.: +38 048 723 80 75
  *E-mail address*: bekshaev@onu.edu.ua (A.Ya. Bekshaev)

Properties of the OV beams produced with the help of such CGHs and transformations they perform to beams already possessing OVs were studied in detail in a series of experimental and theoretical works [32–48]. In particular, it was shown that a diffracted beam of a certain selected diffraction order is mathematically described in the same way as the beam formed after passing a spiral phase plate (helicoidal phase step) (see, e.g., Refs. [49–51]); such paraxial beams constitute the family of "Hypergeometric-Gaussian beams", also called "Kummer beams" [40,43]. However, most results relate to the common case of a thin (plane) HE with EPS. Along with many useful properties (simple fabrication technique, loose alignment requirements, possibility of simultaneous production of many diffracted beams with different singular properties and of suitable regulation of the diffracted beam spatial structure, etc.) they have essential disadvantages: low intensity of any individual diffracted beam, low angular and spectral selectivity and, hence, high sensitivity of the generated OV beam spatial parameters to uncontrollable conditions of the HE-induced transformation.

It is known from the general concepts of holography that a natural way of eliminating these drawbacks is to employ volume (thick) holograms [27–29] based on the 3D interference patterns. Additionally, the 3D interference plays an important role in the techniques of dynamic holography [30] which may be advantageous for the real-time control of the generated OV beams. All mentioned facts make it attractive to consider possibilities of using the volume HE in the OV beam creation and transformation practice, and to analyze the peculiar action of the volume HE with EPS in more detail. However, in the known literature only a few attempts to touch upon these subject were reported [36,37]. The known works have demonstrated reality of the expected advantages of the volume HE with EPS in efficiency but they paid no special attention to peculiar details of the diffracted OV beam formation, its specific spatial properties and how they are interrelated with the HE structure. At the same time, an analogy with the well-studied situation of thin HE suggests that the HE geometry, its position with respect to the incident beam and characteristics of the incident beam itself may affect the diffracted beam parameters, and this can be used for their control and for purposeful generation of OV beams with prescribed properties.

Hence, the problem of systematic investigation of the process of the beam transformation in a HE with EPS is relevant, and this work presents first results of its solution. In Sec. 2, we start with the detailed description of the HE structure as a regular 3D spatially inhomogeneous distribution of the refraction index. For light propagation, the HE is considered as a sequence of thin HE whose action is analyzed in Sec. 3 on the base of the known theory. Sec. 4 is devoted to calculation of the light beam produced by the whole thick HE within the frame of linear single-scattering approach – as a superposition of contributions scattered by separate thin layers. It is finalized by the general formulas for the diffracted beam parameters, which are illustrated by numerical examples in Sec. 5. General conclusions, the author's assessment of the results and probable prospects of further development are briefly outlined in Sec. 6.

## 2. The hologram structure

In contrast to the case of thin CGH whose action can be fairly understood basing exclusively on the spatial periodicity of the equivalent transparency, and the beam transformation can be studied from the general positions common to all planar gratings with EPS [39–42], in analysis of the volume HE it will be more suitable to employ an explicit reference to certain specific hologram structure with proper allowance for the procedure of its preparation. In this work, we consider a rather general model of the hologram recording. In this process, two waves take part (Fig. 1): the reference wave with regular wavefront,

$$E_r(x_r, y_r, z_r) = u_r(x_r, y_r, z_r)\exp(ikz_r), \tag{1}$$

and the subject wave in the form of a circular OV beam [1–5]

$$E_s(x_s, y_s, z_s) = u_s(x_s, y_s, z_s)\exp\left[ikz_s + il\arctan\left(\frac{y_s}{x_s}\right)\right]. \tag{2}$$

In Eqs. (1) and (2), both waves possess the same wavenumber $k$ (the radiation wavelength is $\lambda = 2\pi/k$) and are paraxial in the accompanying coordinate frames $(x_r, y_r, z_r)$ and $(x_s, y_s, z_s)$ (see Fig. 1), which are coupled with the "absolute" frame $(x, y, z)$ by relations

$$x_j = x\cos\theta_j - z\sin\theta_j, \quad z_j = x\sin\theta_j + z\cos\theta_j, \quad y_j = y \quad (j = r, s). \tag{3}$$

Functions $u_r(x_r, y_r, z_r)$ and $u_s(x_s, y_s, z_s)$ are slowly varying in the wavelength scale; for example, in the most common practical case where the reference beam is Gaussian and the subject wave is a circular Laguerre-Gaussian mode with an $l$-charged OV [3,52],

$$u_r(x_r, y_r, z_r) = A_r \frac{1}{b_r}\exp\left(-\frac{x_r^2 + y_r^2}{2b_r^2} + ik\frac{x_r^2 + y_r^2}{2R_r} - i\arctan\frac{z_r - z_{r0}}{kb_r^2}\right), \tag{4}$$

$$u_s(x_s, y_s, z_s) = A_s \frac{1}{b_s}\left(\frac{\sqrt{x_s^2 + y_s^2}}{b_s}\right)^{|l|}\exp\left(-\frac{x_s^2 + y_s^2}{2b_s^2} + ik\frac{x_s^2 + y_s^2}{2R_s} - i(|l|+1)\arctan\frac{z_s - z_{s0}}{kb_s^2}\right). \tag{5}$$

Here $b_j \gg \lambda$ and $R_j \gg \lambda$ ($j = r, s$) are the beam transverse radii and the wavefront curvature radii, $z_{j0}$ are positions of the waist cross sections of the involved beams with amplitude coefficients $A_j$. In most cases, the longitudinal variations of the beams within the hologram depth are negligible, and then $b_j$ and $R_j$ are constant while the last terms in parentheses of Eqs. (4) and (5) can be omitted (included into $A_j$).

The absolute frame is attached to the recording medium that fills a plane-parallel layer, concluded between the coordinate planes $z = \pm d$. The interference pattern formed in the medium is described by function

$$I(x, y, z) = |E_r(x_r, y_r, z_r) + E_s(x_s, y_s, z_s)|^2 = |u_r|^2 + |u_s|^2 + u_r^* u_s \exp(i\Phi) + u_r u_s^* \exp(-i\Phi), \tag{6}$$

where

$$\Phi \equiv \Phi(x, y, z) = k(z_s - z_r) + il\arctan\frac{y_s}{x_s} = qx + pz + il\arctan\left(\frac{y}{x\cos\theta_s - z\sin\theta_s}\right), \tag{7}$$

$$q = k(\sin\theta_s - \sin\theta_r), \quad p = k(\cos\theta_s - \cos\theta_r). \tag{8}$$

In the recording medium, this intensity distribution creates, in general case, inhomogeneity of the (complex) refraction index $n(x, y, z)$. Under conditions of linear response [27]

$$n(x, y, z) = n_0 + \Delta n(x, y, z) = n_0\left[1 + \gamma_p I(x, y, z)\right], \tag{9}$$

coefficient $\gamma_p$ allows for sensitivity of the medium refraction and absorption to the light exposition. When a readout wave with complex amplitude $E(x, y, z)$ falls onto the medium, it is scattered by the inhomogeneity (9). In general, each point $(x, y, z)$ of the medium can be considered as a source of a scattered wave with initial complex amplitude proportional to $E(x,y,z)\Delta n(x,y,z)$; in arbitrary point $P_2$ behind the hologram, the field $E_2(P_2)$ is formed by joint action of all these sources. Calculation of this field is a difficult problem that would usually be solved via certain approximate procedures [27,28,53]; to demonstrate the special features associated with the HE volume nature, we restrict the present consideration to the simple situation of single scattering and the first Born approximation [54]. In this case, one neglects the incident field variations upon passing the hologram depth, and the whole HE can be treated as a sequence of thin layers parallel to its input face [27]. This approach is used in this paper as it permits employing the well known results obtained in the theory of thin HE with EPS [40–42]: The resultant diffracted field appears as a superposition of secondary waves independently formed as a result of the incident wave diffraction by the consecutive thin layers of the recording medium.

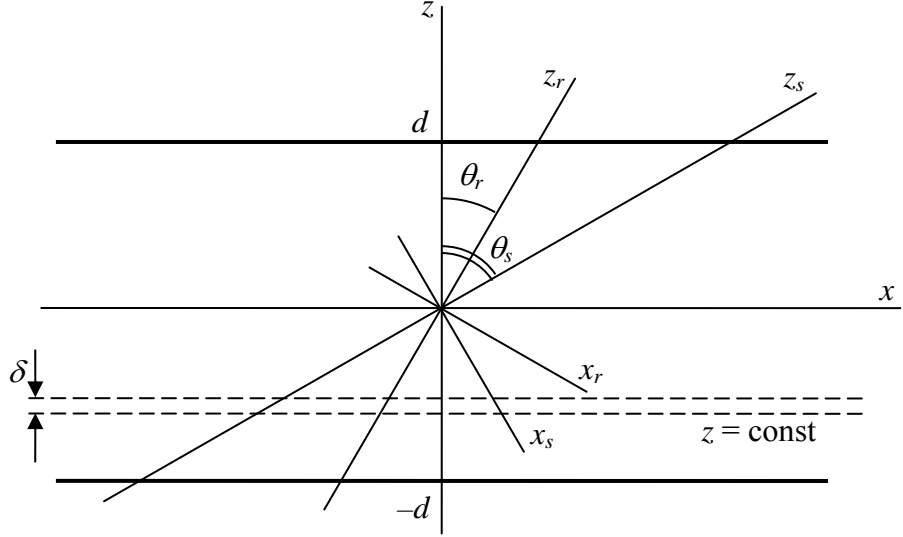

Fig. 1. Formation of the HE in a plane-parallel photosensitive layer of thickness 2*d*. The reference wave axis is $z_r$, the subject wave is directed along axis $z_s$. Both waves are described in their proper coordinate frames $(x_r, y_r, z_r)$ and $(x_s, y_s, z_s)$, frame $(x, y, z)$ is attached to the HE; axes $y_r$, $y_s$ and $y$ coincide and are normal to the figure plane. A selected thin layer of thickness $\delta$ is shown by dashed lines.

### 3. Diffraction in a thin hologram layer

Now we consider a thin layer of the medium with recorded interference pattern, situated near the current plane $z$ = const and with width $\delta \ll \lambda$ (see Fig. 1). Spatial dependence of the layer transmittance is determined in accordance with (9):

$$T(x,y,z) = \exp(ikn\delta) = \text{const} \cdot \exp\left[ik\delta n_0 \gamma_p I(x,y,z)\right]. \tag{10}$$

If a wave with the spatial field distribution $E_1(x,y,z)$ falls onto this layer, after its passage the field is modulated,

$$E_1(x,y)T(x,y,z) \tag{11}$$

where $T(x,y,z)$ follows from Eqs. (6) – (10). In general, functions $u_r$ and $u_s$ change relatively slowly compared to $\Phi$, and their variations can be neglected. Then, due to Eqs. (10) and (6), $T(x,y,z)$ appears to be a periodic function of $\Phi$. Consequently, it can be represented by a

Fourier series, $T(x,y,z) = \sum_{n=-\infty}^{\infty} T_n \exp[im\Phi(x,y,z)]$, and the transmitted beam amplitude (11) is expressed by the sum

$$E_1(x,y,z)T(x,y,z) = \sum_{n=-\infty}^{\infty} T_n E_1(x,y,z)\exp[im\Phi(x,y,z)], \qquad (12)$$

where each summand corresponds to a certain diffraction order. Now we shall separately consider a single diffracted beam of the order $m$, for which the quantity $E_1(x,y,z)\exp[im\Phi(x,y,z)]$ constitutes an initial field distribution in plane $z$ = const. In an arbitrary point $P_2$ behind this layer, the diffracted field of this order is determined via the Kirchhoff formula [41]

$$E_2(P_2;z) = \frac{k}{2\pi i} \int E_1(x,y,z) \exp[im\Phi(x,y,z)] \frac{e^{ik|PP_2|}}{|PP_2|} \frac{1+\cos(z,\overrightarrow{PP_2})}{2} dxdy \qquad (13)$$

where $|PP_2|$ is the distance between $P_2$ and the point $P$ in the considered layer (see Fig. 2a, b) with coordinates $(x, y, z)$, $(z, \overrightarrow{PP_2})$ is the angle between axis $z$ and vector $\overrightarrow{PP_2}$ (see Fig. 2b), and integration is performed over the whole plane $z$ = const. Eq. (13) provides a good approximation if $|PP_2|$ exceeds several wavelengths.

Let us consider an incident wave in the form of a paraxial beam approaching the HE at an angle $\theta_1$,

$$E_1(x_1,y_1,z_1) = u_1(x_1,y_1,z_1)\exp(ikz_1) \qquad (14)$$

In plane $z$ = const its field is expressed by

$$E_1(x,y,z) = u_1(x\cos\theta_1 - z\sin\theta_1, y, x\sin\theta_1 + z\cos\theta_1)\exp[ik(x\sin\theta_1 + z\cos\theta_1)] \qquad (15)$$

(the coordinate transformations similar to Eqs. (3) were used for transition between frames $(x_1, y_1, z_1)$ and $(x, y, z)$). It is the expression (15) that is to be substituted into the Kirchhoff integral (13).

As is well known [39–42], the field diffracted by the layer is confined near the fixed direction; we identify it with axis $z_2$ (see Fig. 2) and introduce the accompanying frame $(x_2, y_2, z_2)$. Then the diffracted field can be described as a paraxial beam that propagates along axis $z_2$:

$$E_2(P;z) \equiv E_2(x_2,y_2,z_2;z) = u_2(x_2,y_2,z_2;z)\exp(ikz_2); \qquad (16)$$

plane $z_2$ = const is the transverse cross section of the diffracted beam (TP$_2$ in Fig. 2). After the substitution of expressions (14) and (16) into Eq. (13), with allowance for (7) and for the coordinate transformations, one arrives at

$$u_2(x_2,y_2,z_2;z) = \frac{k}{2\pi i} \int u_1(x\cos\theta_1 - z\sin\theta_1, y, x\sin\theta_1 + z\cos\theta_1) \frac{1+\cos(z,\overrightarrow{PP_2})}{2|PP_2|} \times$$

$$\times \exp\left\{ik\left(x\sin\theta_1 + z\cos\theta_1 + |PP_2| - z_2\right) + im\left[qx + pz + l\arctan\left(\frac{y}{x\cos\theta_s - z\sin\theta_s}\right)\right]\right\} dxdy. \qquad (17)$$

In usual conditions, $z_2$ is much more than other coordinates $x$, $y$, $z$, $x_2$ and $y_2$. This fact enables further simplifications of Eq. (17) with the help of Fig. 2: first,

$$|PP_2| = \sqrt{(x_2 - x\cos\theta_2 + z\sin\theta_2)^2 + (y_2 - y)^2 + (z_2 - x\sin\theta_2 - z\cos\theta_2)^2}$$

$$\approx z_2 - x\sin\theta_2 - z\cos\theta_2 + \frac{1}{2z_2}\left[(x_2 - x\cos\theta_2 + z\sin\theta_2)^2 + (y_2 - y)^2\right];$$

second, in the integrand denominator one may accept $|PP_2| \approx z_2$; additionally,

$$\cos\left(z, \overrightarrow{PP_2}\right) = \frac{-x_2 \sin\theta_2 + z_2 \cos\theta_2 - z}{|PP_2|} \approx \cos\theta_2,$$

and, in view of slow variations of the complex amplitude $u_1(x_1, y_1, z_1)$ upon change of the longitudinal coordinate $z_1$,

$$u_1\left(x\cos\theta_1 - z\sin\theta_1, y, x\sin\theta_1 + z\cos\theta_1\right) \approx u_1\left(x\cos\theta_1 - z\sin\theta_1, y, 0\right). \tag{18}$$

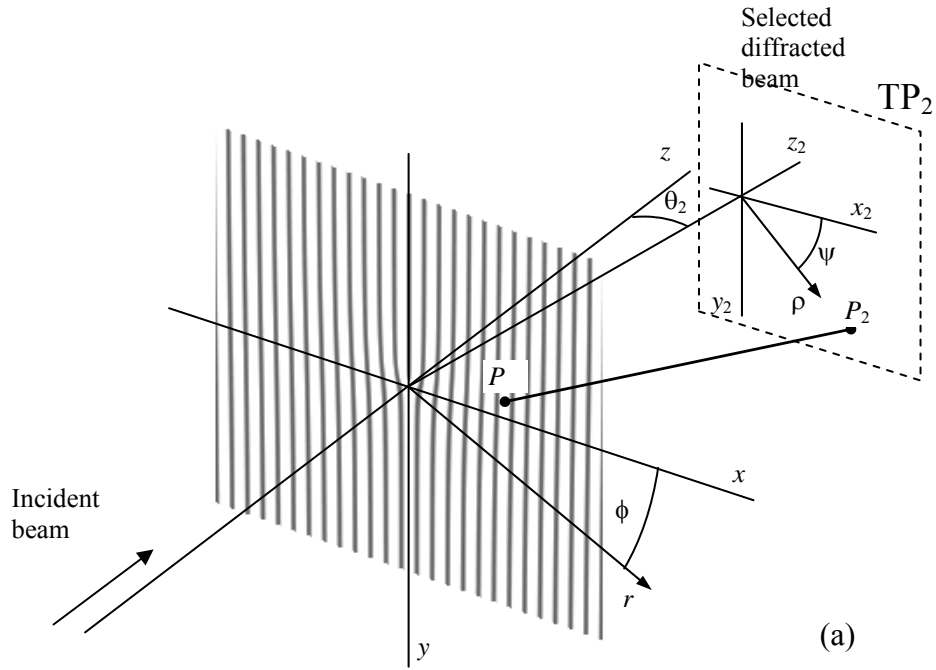

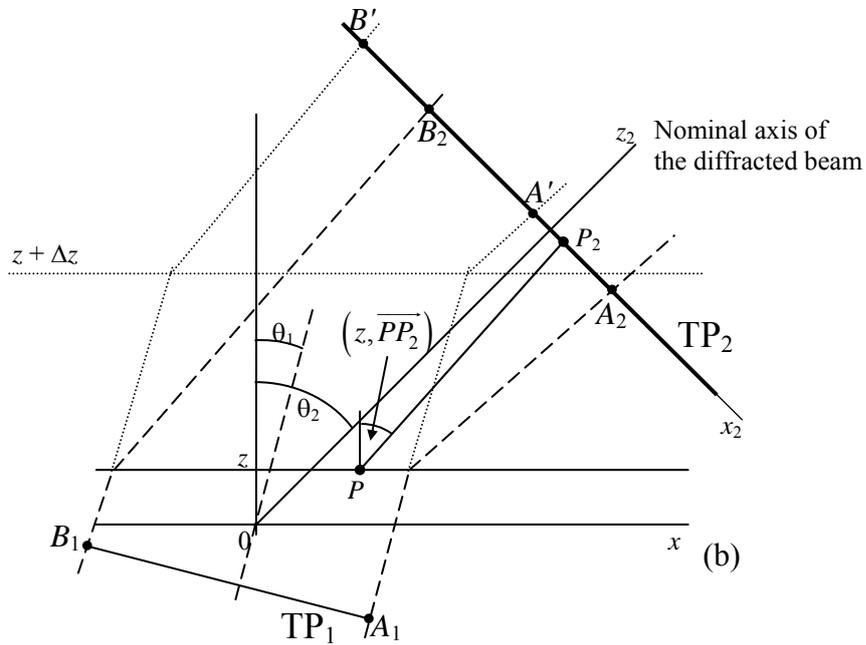

Fig. 2. Formation of a diffracted OV beam in a thin HE (single layer of the volume HE with EPS). $TP_1$ is the transverse plane (cross section) of the incident beam, $TP_2$ is the current cross section of the diffracted beam. Further explanations see in text.

Then Eq. (17) is reduced to the approximate form

$$u_2(x_2, y_2, z_2; z) = \frac{k}{2\pi i z_2} \frac{1+\cos\theta_2}{2} \int u_1(x\cos\theta_1 - z\sin\theta_1, y, 0) \exp\left[iml \arctan\left(\frac{y}{x\cos\theta_s - z\sin\theta_s}\right)\right] \times$$

$$\times \exp\{ix[k(\sin\theta_1 - \sin\theta_2) + mq] + iz[k(\cos\theta_1 - \cos\theta_2) + mp]\}$$

$$\times \exp\left\{\frac{ik}{2z_2}\left[(x_2 - x\cos\theta_2 + z\sin\theta_2)^2 + (y_2 - y)^2\right]\right\} dx dy. \quad (19)$$

In general, terms in the second line of this equation are quickly oscillating multipliers able to strongly diminish the whole integral magnitude up to its vanishing (destructive interference). On the contrary, conditions of constructive interference are realized when

$$(\sin\theta_1 - \sin\theta_2) - m(\sin\theta_r - \sin\theta_s) = 0, \quad (\cos\theta_1 - \cos\theta_2) - m(\cos\theta_r - \cos\theta_s) = 0. \quad (20)$$

Obviously, they hold if

$$m = 1, \theta_1 = \theta_r \text{ and } \theta_2 = \theta_s. \quad (21)$$

In what follows, we will accept these restrictions, and then the second line of Eq. (19) disappears; if, additionally, substitution

$$x \to x\cos\theta_2 - z\sin\theta_2 \quad (22)$$

is applied to the integrand, finally we obtain

$$u_2(x_2, y_2, z_2; z) = \frac{k}{2\pi i z_2} \frac{1+\cos\theta_2}{2\cos\theta_2}$$

$$\times \int u_1(\alpha x + \gamma z, y, 0) e^{il\phi} \exp\left\{\frac{ik}{2z_2}\left[(x_2 - x)^2 + (y_2 - y)^2\right]\right\} dx dy \quad (23)$$

where $\phi = \arctan\left(\frac{y}{x}\right)$ is the polar angle in the coordinate plane $(x, y)$, $\alpha = \frac{\cos\theta_1}{\cos\theta_2}$, $\gamma = \tan\theta_2 \cos\theta_1 - \sin\theta_1$.

In general case, due to the 3D nature of the interference pattern, separate layers of the HE are differently situated with respect to the incident beam. Consequently, expression (23) should be generalized to include "misaligned" situations where the incident beam is shifted from its nominal position; simultaneously, this will allow taking into account the typical practical conditions of the OV beam generation where the incident beam and the HE slightly mismatch because of inevitable arrangement errors. Following to Ref. [42], the incident beam shifted by $s_x$, $s_y$ and inclined by small angles $g_x$, $g_y$ can be expressed in the integrand of (23) by replacement

$$u_1(\alpha x + \gamma z, y, 0) \to u_1(\alpha x + \gamma z - s_x, y - s_y, 0) \exp\{ik[g_x(\alpha x + \gamma z - s_x) + g_y(y - s_y)]\}. \quad (24)$$

Then, after the coordinate substitution

$$x_2' = x_2 - \alpha g_x z_2, \quad y_2' = y_2 - g_y z_2, \quad (25)$$

the output beam profile is determined by equality

$$u_2(x_2, y_2, z_2; z) \equiv u_2^s(x_2, y_2, z_2; z) =$$

$$= \exp\left\{ik\frac{z_2}{2}(\alpha^2 g_x^2 + g_y^2) + ik[g_x(\alpha x_2' + \gamma z - s_x) + g_y(y_2' - s_y)]\right\} u_2^0(x_2, y_2, z_2; z), \quad (26)$$

where

$$u_2^0(x_2, y_2, z_2; z) =$$

$$= \frac{k}{2\pi i z_2} \frac{1+\cos\theta_2}{2\cos\theta_2} \int u_1(\alpha x + \gamma z - s_x, y - s_y, 0) e^{il\phi} \exp\left\{\frac{ik}{2z_2}\left[(x_2' - x)^2 + (y_2' - y)^2\right]\right\} dx dy. \quad (27)$$

## 4. Action of a thick hologram

In accordance with what was said in the end of Sec. 2, the beam diffracted by the whole HE with EPS is formed by the sum of contributions produced by all layers located between $z = -d$ and $z = +d$. In the single-scattering approximation and neglecting the incident beam transformation upon scattering at the preceding layers (each layer "sees" the same incident beam whose profile changes only due to the free-propagation geometric transformation allowed for by relations (18)), its complex amplitude is proportional to the integral

$$u_2(x_2, y_2, z_2) = \int_{-d}^{d} u_2^s(x_2, y_2, z_2; z) dz. \tag{28}$$

Regarding the incident beam arrangement, the integrand function in this relation can be taken in any form of Eqs. (17), (19), (23) or (26).

Equations (26) and (27) constitute the ground for further analysis and/or numerical evaluations. Consider the most usual situation where the incident readout beam is Gaussian and approaches the HE just at its waist cross section, where the wavefront is plane:

$$u_1(x_1 - s_x, y - s_y, 0) = \exp\left[-\frac{(x_1 - s_x)^2 + (y_1 - s_y)^2}{2b^2}\right] \tag{29}$$

($b$ is the beam radius); then the input function in the integrand of (27) appears in form

$$u_1(\alpha x + \gamma z - s_x, y - s_y, 0) = \exp\left[-\left(\frac{\alpha x + \gamma z - s_x}{b\sqrt{2}}\right)^2 - \left(\frac{y - s_y}{b\sqrt{2}}\right)^2\right]. \tag{30}$$

This expression admits partial analytical simplification of the chain of equations (26) – (28): integration over $z$ (see Eq. (28)) can be performed in a closed form. Putting together all terms of Eqs. (26) and (27) with dependence on $z$, one obtains

$$\int_{-d}^{d} \exp\left[-\left(\frac{\alpha x + \gamma z - s_x}{b\sqrt{2}}\right)^2 + ikg_x \gamma z\right] dz = \sqrt{\frac{\pi}{2}} \frac{b}{\gamma} \exp\left[-\frac{1}{2}(kbg_x)^2 - ikg_x(\alpha x - s_x)\right] D(d, x, s_x) \tag{31}$$

where

$$D(d, x, s_x) = \left[\text{erf}\left(\frac{\alpha x + \gamma d - s_x - ikg_x b^2}{b\sqrt{2}}\right) - \text{erf}\left(\frac{\alpha x - \gamma d - s_x - ikg_x b^2}{b\sqrt{2}}\right)\right], \tag{32}$$

and $\text{erf}(\tau) = \frac{2}{\sqrt{\pi}} \int_0^{\tau} e^{-t^2} dt$ is the error function [55]. Note that, in fact, function (32) depends on the four dimensionless variables

$$\bar{x} = x/b, \ \bar{d} = d/b, \ \bar{s}_x = s_x/b, \ \bar{g}_x = kbg_x \tag{33}$$

(all spatial variables are expressed in units of $b$, the angular deviation is measured in units of the incident beam divergence angle [5,52] $(kb)^{-1}$).

Then formulae (26) and (27) yield the final result in the form including only integration over the transverse coordinates. As a result, for the incident Gaussian beam of (29) and for the HE thickness $2d$, the following representation of the diffracted field can be obtained:

$$u_2(x_2, y_2, z_2) = \frac{k}{2\pi i z_2} \frac{1 + \cos\theta_2}{2\cos\theta_2} \sqrt{\frac{\pi}{2}} \frac{b}{\gamma} \times$$

$$\times \exp\left\{-\frac{1}{2}(kbg_x)^2 + ik\frac{z_2}{2}(\alpha^2 g_x^2 + g_y^2) + ik\left[g_x \alpha x_2' + g_y(y_2' - s_y)\right]\right\} \times$$

$$\times \int \exp(-ikg_x \alpha x) D(d, x, s_x) \exp\left[-\left(\frac{y-s_y}{b\sqrt{2}}\right)^2\right] e^{il\phi} \exp\left\{\frac{ik}{2z_2}\left[(x_2'-x)^2 + (y_2'-y)^2\right]\right\} dxdy. \quad (34)$$

Formula (34) can be transformed to more suitable form by employing the expanded set of dimensionless variables (33); after it is supplemented by

$$\bar{x}_2 = x_2/b,\ \bar{y}_2 = y_2/b,\ \bar{y} = y/b,\ \bar{s}_y = s_y/b,\ \bar{g}_y = kbg_y,\ \bar{z}_2 = z_2/kb^2 \quad (33\text{a})$$

(the propagation distance is expressed in units of the Rayleigh range $z_R = kb^2$ [3,5] of the incident beam), one can easily arrive at

$$u_2(\bar{x}_2, \bar{y}_2, \bar{z}_2) = \text{const} \cdot \frac{1}{2\pi i \bar{z}_2} \frac{1+\cos\theta_2}{2\gamma \cos\theta_2} \times$$

$$\times \exp\left\{-\frac{1}{2}\bar{g}_x^2 + i\frac{\bar{z}_2}{2}\left(\alpha^2 \bar{g}_x^2 + \bar{g}_y^2\right) + i\left[\bar{g}_x \alpha \bar{x}_2' + \bar{g}_y(\bar{y}_2' - \bar{s}_y)\right]\right\} \times$$

$$\times \int \exp(-i\alpha \bar{g}_x \bar{x}) D(\bar{d}, \bar{x}, \bar{s}_x) \exp\left[-\left(\frac{\bar{y}-\bar{s}_y}{\sqrt{2}}\right)^2\right] e^{il\phi} \exp\left\{\frac{i}{2\bar{z}_2}\left[(\bar{x}_2'-\bar{x})^2 + (\bar{y}_2'-\bar{y})^2\right]\right\} d\bar{x}d\bar{y} \quad (34\text{a})$$

which constitutes a convenient basis for further calculations and numerical simulations.

## 5. Numerical examples and discussion

Eqs. (26) – (28) and (29) – (34a) supply an efficient means for investigation of the light beams generated by thick HE with EPS when the incident beam is Gaussian. However, possibilities of analytical examination with deriving explicit formulas are rather limited, so in further consideration, numerical approaches will be preferable. In the examples below, we suppose that the incident Gaussian beam crosses the HE input plane exactly in the waist cross section where its complex amplitude is described by Eq. (29); the hologram structure and the recording and readout angles are characterized by the diffraction conditions

$$m = 1,\ \theta_1 = \theta_r = 0\ \text{and}\ \theta_2 = \theta_s = \pi/4 \quad (35)$$

(see Figs. 1, 2 and Eq. (21)). In this paper, the incident beam displacement with respect to the hologram centre is supposed to be absent, $s_x = s_y = 0$.

*5.1. Angular selectivity*

The first specific feature of thick HE with EPS is the quite expected [27–29,36] selectivity of the OV-generation process to angular and spectral deviations from the nominal readout conditions. Dealing with monochromatic input beams, we restrict our analysis to the angular selectivity, which is measured by decrease of the diffracted beam power when the readout beam is slightly inclined with respect to the presumed normal incidence (angles $g_x$ and $g_y$ differ from zero). In the considered approximation, small inclinations oriented along the HE grooves (non-zero $g_y$) are not essential for the output beam parameters; the influence of deviations occurring in the ($xz$) plane is of much more importance.

The effect of angular selectivity originates from the interference between waves diffracted by different layers of the volume HE. In the nominal geometry, where equations (17) – (23) hold, contributions of separate layers $z$ = const in Eq. (28) are in-phase (small phase differences can appear only because of the term $u_1(\alpha x + \gamma z, y, 0)$ in Eq. (23), but the corresponding dependence on $z$ is usually rather weak). In presence of angular deviations, contributions of layers with different $z$ are naturally distinguished by phases, which is reflected by terms $ikg_x\gamma z$ in the exponent of (26) and in the integrand exponent of Eq. (31). Their superposition plays the most important role in the $g_x$-dependence of the total diffracted beam magnitude, which can be roughly estimated, neglecting other contributions of Eq. (31), as

$$\int_{-d}^{d} \exp(ikg_x\gamma z)dz = \frac{2\sin(kg_x\gamma d)}{kg_x\gamma}. \tag{36}$$

In particular, this equation dictates that the interference is destructive and the diffracted beam radiation is suppressed if $g_x = g_a$, where the quantity

$$g_a = \frac{\pi}{k\gamma d} \tag{37}$$

can be taken as a measure of the angular selectivity. For example, under the accepted conditions (35) and $\gamma = 1$, $g_a \approx 0.314 d^{-1}$ rad if $d$ is measured in micrometers. However, quantity (37) corresponds to rather strong suppression of the HE efficiency (although in reality, due to the influence of the integrand terms of (31), discarded in (36), the diffracted beam may not vanish at $g_x = g_a$ completely, yet its magnitude is much less than it could be meaningful in practice). More suitable characteristic is the "half of the HE angular bandwidth" (HBW) – the angular deviation $g_x = g_h$ of the readout wave at which the diffracted beam power equals to half the maximum observed at the nominal incidence $g_x = 0$. Within the studied range of angular misalignments (at least while $g_x < g_a$), the diffracted beam intensity profile shows no noticeable modifications, and variations of the beam power can be traced by the radiation intensity in any point of its cross section. Then the HBW can be estimated via the Taylor expansion of expression (36) due to which the inclination-induced decrease of the beam intensity $I(g_x)$ relative to the maximum $I(0)$ is

$$\frac{I(g_x)}{I(0)} = \left|\frac{u(g_x)}{u(0)}\right|^2 \approx 1 - \frac{1}{3}(kg_x\gamma d)^2 = 1 - \frac{1}{2}\left(\frac{g_x}{g_h}\right)^2 \tag{38}$$

whence

$$g_h = \frac{\sqrt{1.5}}{k\gamma d} \approx \frac{1.22}{k\gamma d} \approx 0.39 g_a. \tag{39}$$

The angular dependence of the diffracted beam power is illustrated by Fig. 3a. The calculated points (asterisks) with rather high accuracy match parabolic curves plotted in accordance with Eq. (38); however, numerically evaluated HBW $g_h$ appears to be slightly less than the analytically derived result (39). This discrepancy is not surprising in view of approximate character of Eqs. (36) and (38). As is predicted by Eqs. (37), (39) and in agreement with the known data on thick the hologram behavior [27–29], the inverse HBW value $g_h^{-1}$ shows approximately linear growth with the HE thickness $2d$ which, however, slightly slows down at high $d$ (Fig. 3b).

*5.2. Evolution of the diffracted beam transverse profile*

The spatial configuration of an OV beam generated by the HE with EPS and laws of its evolution constitute the main interest associated with the study of thick holograms designed for the OV beam generation. Due to the noticed independence of the beam intensity profile on the angular deviations in the meaningful region $g_x < g_a$ (see the paragraph below Eq. (37)), in this subsection we may accept the condition $g_x = 0$.

Typical results presented in Fig. 4 indicate remarkable details concerning the specific role of the HE volume nature. By confrontation of the 1st and 2nd rows of Fig. 4 one can conclude that increasing the HE thickness up to a rather high value of $2d \approx b$ makes practically no special influence on the diffracted beam spatial profile and its evolution; visually one could hardly differentiate such a beam from the OV beam produced by a thin CGH (we let aside the issues of efficiency, selectivity, elimination of multiple diffraction orders, etc. [27,29,36]). As in the well known situations of thin HE with EPS [41,45,49], early stages of the diffracted beam evolution feature the ripple structure owing to the edge wave originating from the HE singularity (the groove bifurcation point [34,35,45]), see 1st column of the 1st and 2nd rows of Fig. 4. Since the readout conditions (35) correspond to the high-angle diffraction [45], both in a thin and in a thick HE, simultaneously with the OV formation, the beam is squeezed in the diffraction plane, and this deformation brings about the practically identical symmetry-breakdown consequences [45,56,57]: the expected bright ring is replaced by the oval with inhomogeneous brightness, and the whole beam pattern rotates in accordance with the transverse energy circulation in the OV generated.

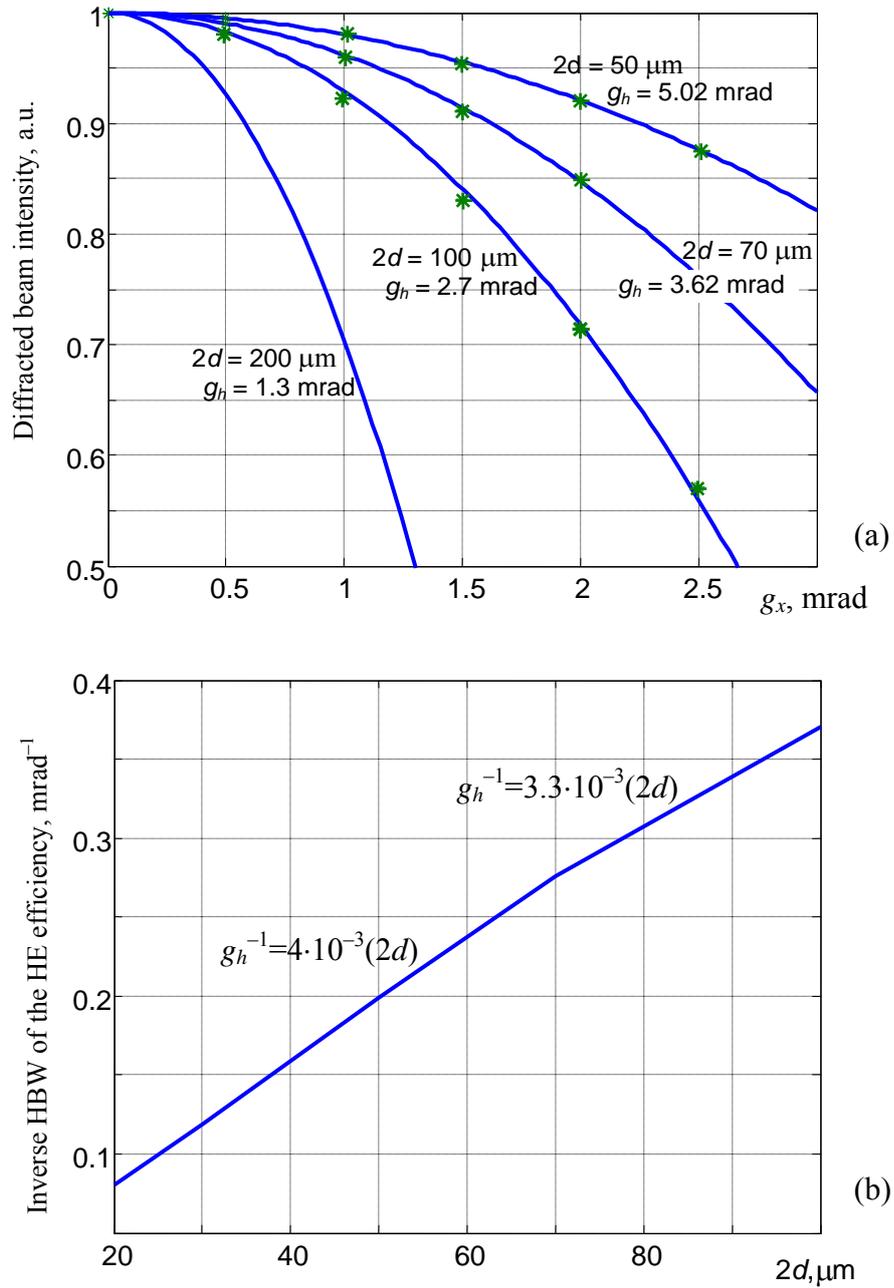

Fig. 3. (a) Angular dependence of the diffracted beam power for $\gamma = 1$ and conditions (35): asterisks denote points calculated by formula (34), curves express approximation (38), (39); (b) Inverse angular HBW of the HE efficiency curve vs the HE thickness: current forms of the proportionality are indicated near the line segments.

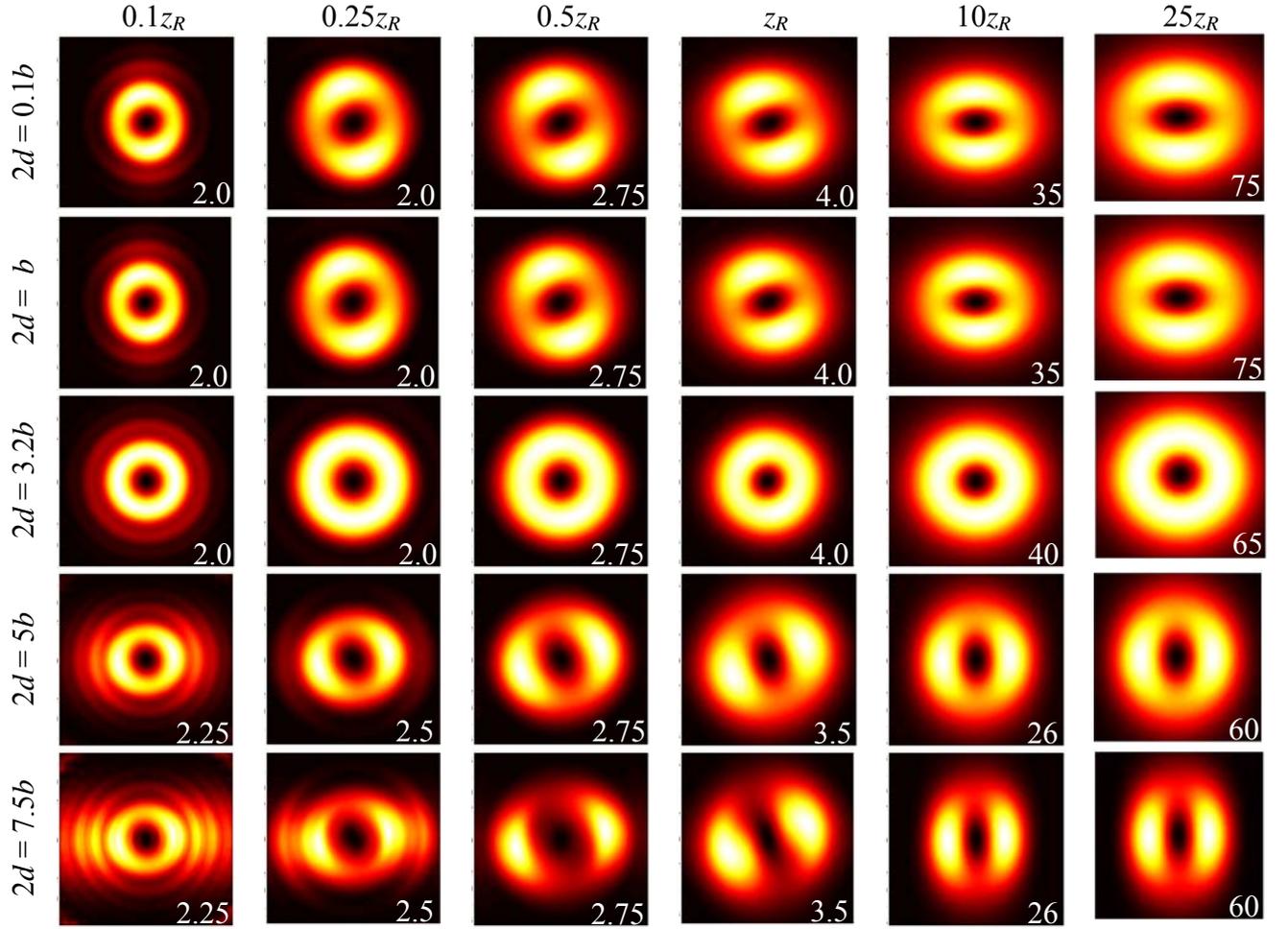

Fig. 4. Evolution of the OV beams obtained after diffraction of the Gaussian beam under conditions (35) in the volume HE with EPS calculated by Eq. (34a). The HE thickness is indicated near each row, corresponding propagation distance is shown above each column in units of the incident beam Rayleigh range; the image absolute sizes (lengths of the square sides) are marked in the lower left corners in units of the incident beam radius $b$.

However, with further increase of the HE thickness, the output beam shape and its way of evolution change rather drastically. At $2d = 3.2b$, the symmetry breaking effect of the high-angle diffraction apparently disappears; the output diffracted beam gets circular symmetry and preserves it within the whole propagation range (3rd row of Fig. 4). With further growth of the HE thickness, the output beam squeezing in the diffraction plane, evident in the initial columns of the 1st and 2nd rows, is replaced by obvious stretching in 4th and 5th rows; accordingly, the propagating beam evolves now in accordance with the scheme typical for transformation of an OV beam initially squeezed in the vertical plane [56,57]. In all cases except $2d = 3.2b$ (3rd row of Fig. 4), beams with anisotropic OV are formed, and this anisotropy exists during the whole beam propagation up to the far-field conditions (last column of Fig. 4).

Transition from squeezed to stretched profile of the diffracted beam can be directly linked to the many-layer model of the HE (Fig. 2b). Let the incident beam with diameter $A_1B_1$ is diffracted by the thin layer close to $z$; then the beam with width $A_2B_2 = \eta A_1 B_1 (\cos\theta_2 / \cos\theta_1)$ is produced in the output plane TP$_2$ ($\eta$ is the coefficient close to 2 [41] allowing for the difference between the usual definitions of the Gaussian beam diameter $A_1B_1$ and the OV beam size $A_2B_2$). If the same incident beam diffracts at the thin layer located near $z + \Delta z$, the corresponding output radiation occupies in the plane TP$_2$ another region $A'B'$ transversely shifted by $(\Delta z/\cos\theta_1)\sin(\theta_2 - \theta_1)$. Similarly, the intermediate diffracting layers contribute to intermediate localizations of the diffracted radiation so that the total transverse size of the output beam, diffracted by the whole HE depth between $z$ and $z + \Delta z$, is approximately characterized by the length $A_2B' = A_2B_2 + (\Delta z/\cos\theta_1)\sin(\theta_2 - \theta_1)$. Assuming conventionally $A_1B_1 = 2b$ and $\Delta z = 2d$, we find that the total diffracted beam width $2b_{2x} = A_2B'$ in the diffraction plane can be estimated as

$$2b_{2x} = 2\eta b \frac{\cos\theta_2}{\cos\theta_1} + \frac{2d}{\cos\theta_1}\sin(\theta_2 - \theta_1), \qquad (40)$$

which explicitly shows relations between the diffracted beam size and the HE thickness $2d$. In particular, under conditions (35), Eq. (40) shows that with growing $d$ the output beam size in the diffraction (horizontal) plane $b_{2x}$ varies as

$$b_{2x} = \eta b \cos\theta_2 + d\sin\theta_2 = 0.71(\eta b + d).$$

Hence, it formally follows that the diffracted beam becomes symmetric, $b_{2x} \approx \eta b$, when the HE thickness $2d \approx \eta b$. The deviation from condition $2d \approx 3.2b$ actually observed in 3rd row of Fig. 4 can be ascribed to the complex shape of the beam profiles involved and to the approximate nature of parameters $b$ and $\eta$ as characteristics of the beams' spatial sizes.

Horizontal stretching of the diffracted beam spots in the initial panels of the 4th and 5th rows of Fig. 4 is coupled with the more articulated ripple structure; most probably, the same ripple modulations that are not visible in the upper rows of Fig. 4 because of negligible diffracted beam intensity, become observable when the diffracted beam is elongated in the $x$-direction (4th and 5th rows, 1st and 2nd columns of Fig. 4).

It should be emphasized that this mechanism of the output beam profile formation implies that absolute quantities of the light energy scattered by all layers parallel to the input face of the HE are equal: the incident beam attenuation, e.g., because of diffraction in preceding layers, is negligible. For the volume HEs, this restriction, though realizable in practice, is rather artificial and contradicts to their most expected property of high diffraction efficiency. The simplest and rather rough way to take this attenuation into account is to assume that the incident beam intensity decays homogeneously over the beam cross section in agreement with the Bouguer's law; then the exponential multiplier $\exp[-\kappa(z+d)]$ should be added to the integrand of Eq. (28) where $\kappa$ is the amplitude coefficient of light extinction in the HE depth. Then Eqs. (34), (34a) can still be used for the diffracted beam profile analysis provided that $g_x$ is replaced by $g'_x = g_x + i(\kappa/k\gamma)$ in expressions (31) and (32). Images of Fig. 5 show that even considerably weak attenuation of the incident beam essentially modifies the beam profile obtained in the thick HE, up to disappearance of the OV itself, not to mention the beam stretching in the diffraction plane and all the accompanying effects clearly seen in the lower rows of Fig. 4. For example, at $b = 0.2$ mm, which is typical for the He-Ne lasers widely used for the OV generation, the extinction coefficient compatible with conditions of Fig. 5 is $\kappa = 1$ mm$^{-1}$. Needless to say, one cannot expect to observe any specific beam profile peculiarities associated with the HE thickness if the latter exceeds the extinction length $\kappa^{-1}$.

As a final remark on how the HE thickness influences the diffracted beam profile, we note that the simplified linear theory presented in this work is of limited applicability, and in further developments, a dynamic approach allowing for the continuous energy exchange between the readout and diffracted waves, based, e.g., on the coupling mode theory [28,29], should be employed.

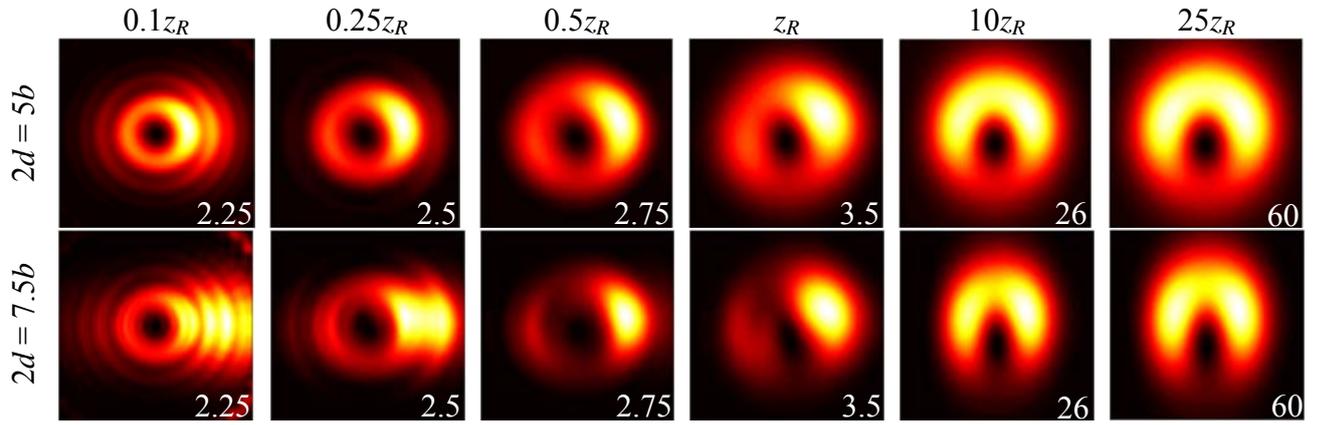

Fig. 5. Evolution of the OV beam profiles calculated similarly to images of Fig. 4 but with non-zero extinction coefficient $\kappa = 0.2/b$ mm$^{-1}$; the system of notations is the same as in Fig. 4. Only results for very thick HE with the most impressive profile distortions are shown (cf. 4$^{th}$ and 5$^{th}$ rows of Fig. 4).

### 5.3. Morphology of the produced OVs

In the previous subsection, our interests were focused on the overall spatial characteristics of the OV beams produced due to diffraction in the thick HE with EPS. Description of the OVs nested in these beams constitutes a separate task that can be accomplished basing on the known methods of studying the OV morphology [58–60]. General description of the OV morphology is grounded on the fact that in the nearest vicinity of an OV axis ($x_2 = y_2 = 0$), the complex amplitude distribution can be represented as

$$u(x_2, y_2) \propto (\xi + \chi) x_2 + i(\xi - \chi) y_2 \qquad (41)$$

where $\zeta$ and $\chi$ are certain complex numbers depending on $z_2$. From this representation, the OV morphology characteristics follow [58,59]: parameter of orientation

$$\phi_a = \frac{i}{2} \ln\left(\frac{\xi \chi^*}{|\xi||\chi|}\right) = \frac{1}{2}(-\arg \xi + \arg \chi) \qquad (42)$$

whose physical meaning is that angle $\phi_a$ determines orientation of the minor axis of the so called anisotropy ellipses (contours of equal amplitude (41) in the ($x_2$, $y_2$) plane, see Fig. 6), and the form-factor of those ellipses (major to minor axes ratio $w_+/w_-$):

$$\frac{w_+}{w_-} = \frac{|\xi| + |\chi|}{|\xi| - |\chi|} \qquad (43)$$

(see also Refs. [60,61]). The wavefront morphology in the OV area is also described by the parameters (42) and (43) [61]: on a round trip near the vortex core, the phase grows most rapidly when crossing the major axis of the anisotropy ellipse and most slowly near its minor axis; the maximum and minimum rates of the phase growth relate as $w_+/w_-$.

The morphology parameters (42) and (43) provide the standard characterization of the OVs generated by the HE [45]. Their evolution in the propagating beams, whose spatial profiles are presented in Fig. 4, is described by Fig. 7; the propagation distance is expressed in units of the Gouy phase (see, e.g. Refs. [3,5]) of the incident Gaussian beam

$$\zeta = \arctan \bar{z}_2 = \arctan\left(\frac{z_2}{kb^2}\right). \qquad (44)$$

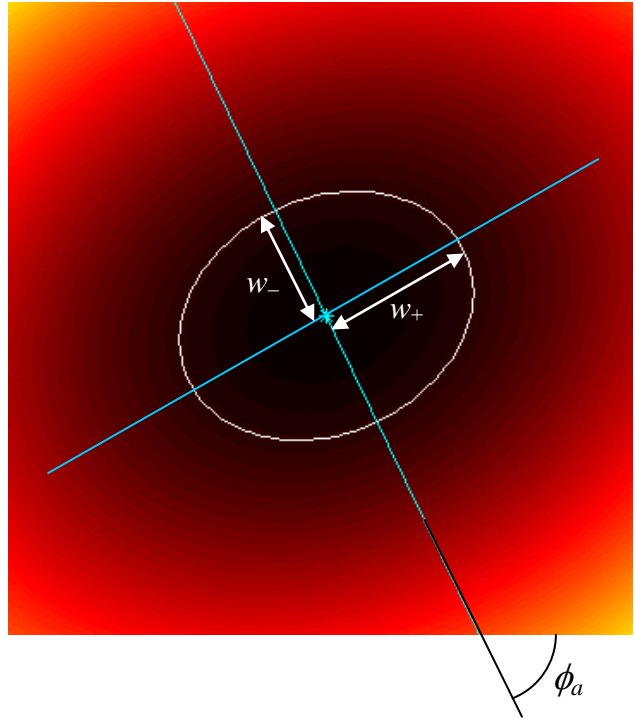

Fig. 6. Near-OV part of the intensity distribution for the diffracted beam presented in 2$^{nd}$ row, 3$^{rd}$ column of Fig. 4 ($z_2 = 0.5z_R$, $2d = b$). The anisotropy ellipse is shown with the morphology parameters of Eqs. (42) and (43).

It is seen that in a wide range of the HE thicknesses (at least, even at $2d = b$), the OV morphology behaves in a fair compliance with the theory of thin CGH under the high-angle diffraction conditions [45]. The form-factor curves marked 0.1 and 1 (Fig. 7a) show only quantitative discrepancies between each other whereas the orientation-angle curves visually almost coincide (see the inset in Fig. 7b for details). Anyway, all curves marked 0.1 and 1 look similar to the prototype results of Ref. [45]; the characteristic oscillations in the near-field region are associated with the ripple structure in the corresponding beam profiles in the left upper corner of Fig. 4. Curve 3.2 of Fig. 7a shows that apparently circular pattern of the 3rd row of Fig. 4 is not perfect, although it presents the best result of our efforts to find conditions for the output beam to be circularly symmetric. However, small deviations from the circular geometry expressed by oscillations of curve 3.2 in Fig. 7a can hardly be meaningful practically. At least, during the numerical analysis (as well as, most probably, in an expected experiment) it was impossible to reliably identify the anisotropy ellipse orientation, for which reason the curve 3.2 is absent in Fig. 7b.

The morphology of OVs obtained in extremely thick HE with EPS (curves marked 5 and 7.5 in Fig. 7) show additional peculiarities associated with the near-field oscillations. Their amplitude drastically grows as well as the range of the propagation distances at which they are observable (up to $z_2 \sim 0.5kb^2$ that corresponds to $\zeta \sim 0.6$). Undoubtedly, this is related to the mechanism of the diffracted beam formation in the process of superposition of many OV beams laterally shifted with respect to each other in the diffraction plane (see Sec. 4). Despite the apparent irregularity of the oscillations' pattern, the explicit correlation between extrema of different curves is obvious; additionally, minima of the form-factor curves 5 and 7.5 in Fig. 7a can be associated with regions of the fastest downfall of the orientation-angle curves with the same names. Interestingly, curve 3.2 of Fig. 7a has maxima where other curves possess minima and vice versa. Taken in a whole, these features testify for the common reason of the oscillatory details of all curves in Fig. 7 which appear due to interference with the edge wave originating from the HE singularity [41,46].

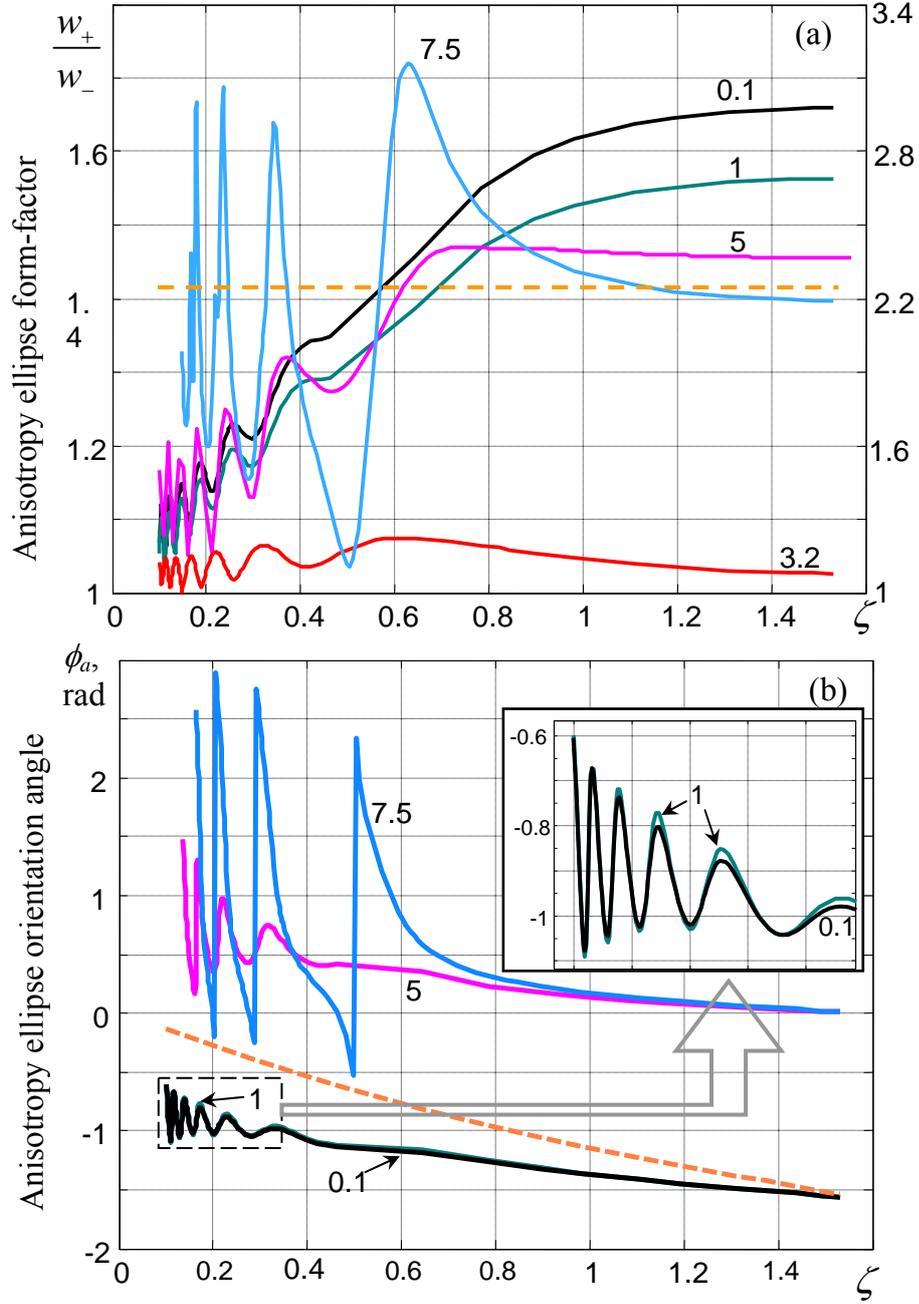

Fig. 7. (a) Form-factor (43) and (b) orientation angle (42) of anisotropy ellipses for the OV beams, presented in Fig. 4, as functions of the Gouy phase parameter (44). Values of the HE thickness $2d$ in units of $b$ are marked near each curve; data for curve 7.5 in panel (a) are given at the right vertical axis. Plot of $\phi_a$ for $2d = 3.2$ is not presented because the anisotropy ellipses are close to circumferences whose orientation is indeterminate (cf. the corresponding form-factor curve in panel (a)); initial regions of curves 1 and 0.1 (dotted rectangle) in panel (b) is magnified in the inset. For comparison, behavior of the morphology parameters in an asymmetrically deformed Laguerre-Gaussian beam with initial geometrical squeezing $b_y/b_x = \cos\theta_2/\cos\theta_1 = 1.41$ (see Fig. 2 and Eq. (35)) is characterized by dashed lines calculated following Ref. [57].

## 6. Conclusion

By this paper, we have launched a systematic study of the OV generation technique employing the volume HEs. A simple theory of a thick HE has been presented that is based on the linear single scattering model and the Born approximation in which the HE is considered as a consequence of thin layers parallel to the hologram surface. This approach enables to consider the volume HE as a generalization of a thin CGH model that was developed and substantiated in the previous works [35–46].

As an example, the OV generation from a Gaussian incident beam with plane wavefront whose axis exactly coincides with the nominal HE axis (regime of perfect alignment) has been analyzed numerically. The proposed approach permitted us to quantitatively describe such specific features of the volume HE with EPS as reduction of the allowed diffraction orders, high angular and spectral selectivity. Besides, peculiar influence of the HE thickness on the output beam intensity profile has been revealed. At low and moderate HE thicknesses (less than the incident beam size), the diffracted beam shape is satisfactorily described by the results of the thin HE theory. With growing HE thickness, the diffracted beam shape changes from squeezed to elongated in the diffraction plane; in intermediate regimes, a circular symmetric profile can be realized regardless of the diffraction angle. The morphology parameters (form-factor and orientation of the equal-intensity ellipses in the nearest vicinity of the axis) of the generated OVs are analyzed, and their evolution in the course of the beam propagation is found to be qualitatively similar to the familiar picture known for the OV beams obtained by means of thin HEs. The main feature, shared both by the thin and thick HEs with EPS, is more or less intensive oscillation of the morphology parameters in the near-field zone; however, quantitative parameters of these oscillations essentially differ for different situations. The morphology parameters' oscillations as well as the ripple structure in the transverse beam profiles (see, e.g., the 1st column of Fig. 4) are associated with the divergent edge wave originating from the HE singularity (the groove bifurcation point) that interferes with the "regular" diffracted field [41,42,45,46].

The main simplification of the model presented in this paper is that the incident beam is supposed to propagate invariably within the whole HE depth – the back influence of the diffracted light on the incident beam is neglected. This restriction apparently contradicts to the expected high diffraction efficiency of the volume HE and should be overcome in further non-linear theory allowing for the continuous energy exchange between the readout and diffracted waves, based, e.g., on the coupling mode theory [28,29]. An attempt to take into account the simplest version of mentioned back influence – exponential decay of the incident beam due to the Bouguer extinction law – has shown the strong action of the incident beam extinction on the spatial characteristics of the diffracted beams obtained in the HE with high thickness (Fig. 5).

In this work, we did not touch upon the phase profiles of the generated OV beams. Preliminary investigations show that, compared to the intensity distributions, the phase profiles show specific variations and are, in general, much more sensitive to the nominal scheme violations (e.g., non-zero incidence angles $g_x$, $g_y$ and misalignments $s_x$, $s_y$) whose influence will be inspected elsewhere.

Finally, we would like to emphasize that the presented way of the diffracted field calculation directly addresses combination of a number of OV beams given by Eqs. (26) (27) that differ mainly by the localization in the output transverse plane $TP_2$ (see Fig. 2b). This leads to suggestion that the methods and results of the present work can be useful in other problems associated with superposition of many OV beams whose axes are regularly shifted with respect to each other in the transverse directions.